\documentclass[aps,prd,preprint,superscriptaddress,tightenlines,nofootinbib]{revtex4}

\usepackage{graphicx}% Include figure files
\usepackage{dcolumn}% Align table columns on decimal point
\usepackage{bm}% bold math

\newcommand{\pipi}{\pi^{+}\pi^{-}}
\newcommand{\KK}{K^{+}K^{-}}
\newcommand{\ppbar}{p\overline{p}}
\newcommand{\hadpair}{h^{+}h^{-}}
\newcommand{\ee}{e^{+}e^{-}}

\newcommand{\leppair}{l^{+}l^{-}}

\newcommand{\eetomm}{\ee \rightarrow m^{+}m^{-}}
\newcommand{\eetopipi}{\ee \rightarrow \pipi}
\newcommand{\eetoKK}{\ee \rightarrow \KK}
\newcommand{\eetoppbar}{\ee \rightarrow \ppbar}
\newcommand{\eetohh}{\ee \rightarrow \hadpair}
\newcommand{\eetoll}{\ee \rightarrow \leppair}

\newcommand{\psip}{\psi(2S)}

\newcommand{\mff}{|F_{m}(s)|}

\begin{document}

%\preprint line(s) will be ignored for PRL/PRD
%\preprint{CLEO Draft 05-49A} % For paper draft CBX YY-NN -> Draft YY-NNA
%\preprint{CLEO CONF YY-NN}   % For conference papers
%\preprint{ICHEP ABSnnn}      % For conference papers
\preprint{CLNS 05-1936}       % for CLNS notes
\preprint{CLEO 05-24}         % for CLNS notes

\title{Precision Measurements of the Timelike Electromagnetic \\ 
Form Factors of Pion, Kaon, and Proton}
% for conference papers (ask CLEOAC for appropriate text)
%\thanks{Submitted to the 31$^{\rm st}$ International Conference on High Energy
%Physics, July 2002, Amsterdam}

%-------- INSERT HERE ------------
% Your author list goes here  REMOVE EVERYTHING to END INSERT and
% replace with your authorlist (ask cleoac).

\author{T.~K.~Pedlar}
\affiliation{Luther College, Decorah, Iowa 52101}
\author{D.~Cronin-Hennessy}
\author{K.~Y.~Gao}
\author{D.~T.~Gong}
\author{J.~Hietala}
\author{Y.~Kubota}
\author{T.~Klein}
\author{B.~W.~Lang}
\author{S.~Z.~Li}
\author{R.~Poling}
\author{A.~W.~Scott}
\author{A.~Smith}
\affiliation{University of Minnesota, Minneapolis, Minnesota 55455}
% ...
%\author{(CLEO Collaboration)} %FOR PRD_SPECIAL_CHANGEME
\collaboration{CLEO Collaboration} %FOR PRL,CLNS
\noaffiliation

%-------- END INSERT ------------

%please hard code the date when you have a final draft and submit to CLEOAC
%\date{\today}
\date{September 30, 2005}

\begin{abstract} 
Using 20.7 pb$^{-1}$ of $e^+e^-$ annihilation data taken at $\sqrt{s}=3.671$ GeV 
with the CLEO--c detector, precision measurements of the electromagnetic form 
factors of the charged pion, charged kaon, and proton have been made for timelike 
momentum transfer of $|Q^2|=13.48$ GeV$^2$ by the reaction $e^+e^-\to h^+h^-$.  
The measurements are the first ever with identified pions and kaons of $|Q^2|>4$ 
GeV$^2$, with the results 
$F_\pi(13.48\;\mathrm{GeV}^2)=0.075\pm0.008(\mathrm{stat})\pm0.005(\mathrm{syst})$ 
and 
$F_K(13.48\;\mathrm{GeV}^2)=0.063\pm0.004(\mathrm{stat})\pm0.001(\mathrm{syst})$.  
The result for the proton, assuming $G^p_E=G^p_M$, is 
$G^p_M(13.48\;\mathrm{GeV}^2)=0.014\pm0.002(\mathrm{stat})\pm0.001(\mathrm{syst})$, 
which is in agreement with earlier results. 
\end{abstract}

\maketitle

Electromagnetic form factors of hadrons are among the most important physical 
observables.  They provide direct insight into the distribution of charges, 
currents, color, and flavor in the hadron.  Form factors for spacelike momentum 
transfers, $Q^2>0$, are determined by elastic scattering of electrons from 
hadrons available as targets.  Form factors for timelike momentum transfers, 
$Q^2<0$, are measured by annihilation $e^+e^-\leftrightarrow h^+h^-$.  They lead 
to insight into the wave function of the hadron in terms of its partonic 
constituents.  In this Letter we report on the first precision measurements for 
the timelike electromagnetic form factors of the pion, kaon, and proton, 
for $|Q^2|=13.48$ GeV$^2$, by means of the reactions
\begin{equation}
e^+e^-\to\pi^+\pi^-,\;K^+K^-,\;\mathrm{and}\;p\bar{p}
\end{equation}

Measurements of the timelike form factors of pion and kaon, $F_\pi$ and $F_K$, 
with identified pions and kaons exist for $|Q^2|\le4.5$ GeV$^2$ 
\cite{timelikepionkaon}.  For larger $|Q^2|$ either only upper limits for 
$F(|Q^2|)$ exist, or the few observed hadron pairs, not separately identified as 
pions or kaons, are divided between the two according to the expectations based 
on the vector dominance model (VDM) to obtain $F_\pi$ and $F_K$ \cite{timelikevdm}. 

Timelike form factors of the proton for $|Q^2|>6$ GeV$^2$ were first measured by 
the Fermilab E760/E835 experiments via the reaction $p\bar{p}\to e^+e^-$ 
\cite{e760e835}.  According to perturbative QCD (pQCD) at large momentum 
transfers the timelike form factors of protons are expected to be nearly equal 
to the spacelike form factors, but the Fermilab measurements found the timelike 
form factors to be nearly twice as large.  In this Letter we provide the first 
independent confirmation of the Fermilab observations.

Theoretical predictions of form factors based on pQCD rely on the validity of 
factorization for sufficiently high momentum transfers, and lead to quark 
counting rules, which predict \cite{quarkcounting} that 
$F(|Q^2|)\propto |Q^2|^{1-n}$, where $n$ is the number of quarks, so that 
$F(|Q^2|)\propto \alpha_S/|Q^2|$ for mesons, and 
$F(|Q^2|)\propto \alpha_S^2/|Q^4|$ for baryons. pQCD also predicts \cite{pQCD} 
that the form factors for the helicity--zero mesons $m=\pi,\;K,\;\rho,\;...$ are 
proportional to the squares of their decay constants so that 
$F_\pi(|Q^2|)/F_K(|Q^2|)=f_\pi^2/f_K^2$, as $|Q^2|\to\infty$.

By providing tests of the above predictions, we expect to shed light on the 
important question of the momentum transfers which are sufficiently large to 
validate the use of pQCD, a question which has been in debate for a long time 
\cite{debateLepage,debateIsgur}.

The timelike form factors of the charged helicity--zero mesons are related 
to the differential and total cross sections for their 
pair production by
\begin{eqnarray}
\frac{d\sigma_{0}(s)}{d\Omega}(\eetomm) = \frac{\alpha^{2}}{8s}~
\beta^{3}_{m}~\mff^{2}\mathrm{sin}^{2}\theta,
\label{mesonad}
\end{eqnarray}
where $m$ = $\pi$ or $K$, $\alpha$ is the fine-structure constant, 
$\beta_{m}$ is the meson velocity in the laboratory system, 
$s=|Q^2|$ is the center-of-mass energy squared, $\mff$ is the electromagnetic 
form factor of the meson, and 
$\theta$ is the laboratory angle between 
the meson and the positron beam.  

The $e^+e^-\to p\bar{p}$ differential cross sections are related to the magnetic 
($G^p_M$) and electric ($G^p_E$) form factors of the proton. With $\tau=4m_p^2/s$
\begin{eqnarray}
\frac{d\sigma_0(s)}{d\Omega} \! =\! \frac{\alpha^2}{4s} \beta_p \!\!\left[ 
|G_M^p(s)|^2(1\! +\!\cos^2\!\theta)\! +\! \tau |G^p_E(s)|^2\sin^2\!\theta\right]
\label{baryonad}
\end{eqnarray}
In the present measurements we do not have sufficient statistics to separate 
$G^p_E(s)$ and $G^p_M(s)$.  We therefore analyze our data with the two 
assumptions, $G^{p}_{E} = G^{P}_{M}$ and $G^{p}_{E} = 0$, as in Ref. \cite{e760e835}. 

The $e^+e^-$ annihilation data samples used in the present measurements consist 
of 20.7 pb$^{-1}$ at $\sqrt{s}=3.671$ GeV and 2.89 pb$^{-1}$ at the $\psi(2S)$ 
mass, $\sqrt{s}=3.686$ GeV. The data were collected at the Cornell Electron 
Storage Ring (CESR) with the detector in the CLEO--c 
configuration \cite{CLEOcDetector}. The detector is cylindrically 
symmetric and provides 93$\%$ coverage of solid 
angle for charged and neutral particle 
identification.  The detector components important for this analysis are 
the wire vertex detector (ZD), 
the main drift chamber (DR), the ring-imaging Cherenkov detector (RICH), 
and the CsI crystal calorimeter (CC). They are operated 
within a 1.0 T magnetic field produced by a superconducting 
solenoid located directly outside of the CC.   
The ZD and DR detect charged particles, and 
the DR provides measurement of their momenta with a precision of
$\sim$0.6$\%$ at $p$ = 1 GeV/c and ionization energy loss ($dE/dx$). 
The RICH detector provides particle identification, 
and covers 80$\%$ of the solid angle.  
The combination of $dE/dx$ and information from the RICH detector allows 
for separating different charged particle species.  
The CC allows precision measurements of 
electromagnetic shower energy and 
position. 

\begin{figure}[!tb]
\hspace*{-0.2in}\includegraphics*[width=4.2in]{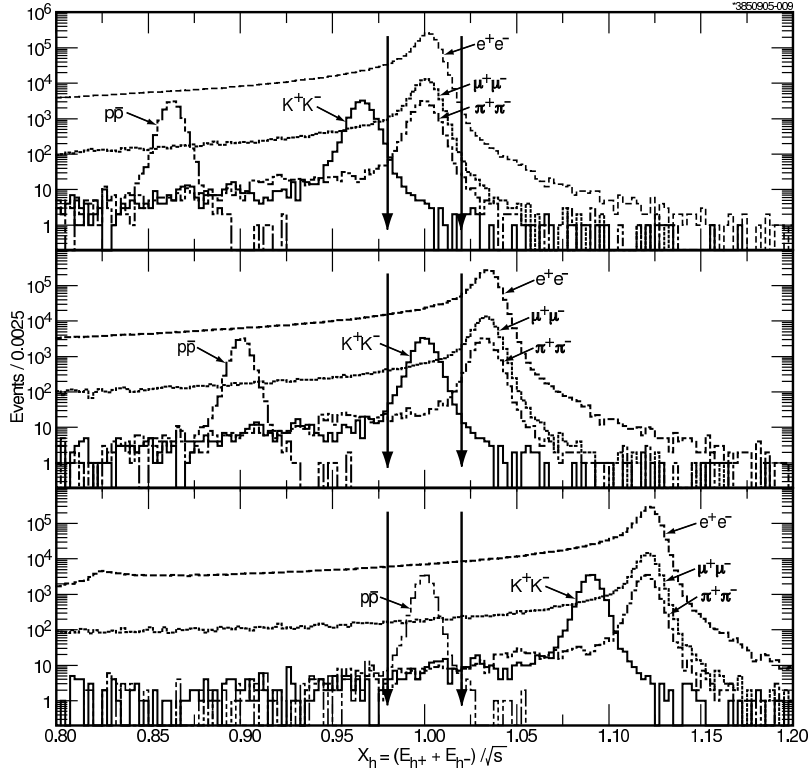}
\caption{MC distributions as a function of $X_{h}$ $\equiv$ 
$(E_{h+} + E_{h-}) / \sqrt{s}$, where $h = \pi$ (top), $h = K$ (middle), 
and $h = p$ (bottom).  The signal regions are defined as 
0.98 $<$ $X_{h}$ $<$ 1.02 and are designated by the vertical arrows.}
\label{fig:xwideinit}
\end{figure}

A fully reconstructed event is required to have two charged particles and zero 
net charge.  The charged particles in the $\pipi$ final state analysis must have 
$|$cos$\theta|$ $<$ 0.75 and have an associated shower in the CC.  The 
charged particles in 
the $\KK$ and $\ppbar$ analyses must have $|\cos\theta|$ $<$ 0.80.  
Each charged particle is required to satisfy standard requirements for 
track quality and distance of closest approach to the interaction point.  
For the $\pipi$ and $\ppbar$ analyses, the two charged particles must 
have a net momentum $<$ 100 MeV/c, and for the $\KK$ analysis,
they must have a net momentum $<$ 60 MeV/c.

\begin{figure}[!tb]
\includegraphics*[width=4.5in]{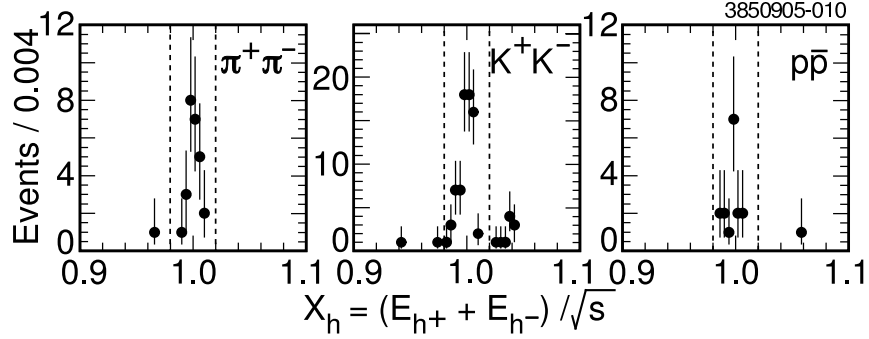}
\caption{Data events as a function of $X_{h}$ for $\pipi$ (left), 
$\KK$ (middle), and $\ppbar$ (right) final states.  The dashed lines 
denote the signal regions defined as 0.98 $<$ $X_{h}$ $<$ 1.02.}
\label{fig:xwidedata}
\end{figure}

The hadronic $\eetohh$ processes, where $h = \pi, K, p$, are studied 
by generating signal Monte Carlo (MC) samples, using GEANT 
\cite{GEANTMC} to simulate the ~\mbox{CLEO--c} detector.  MC samples 
of the leptonic background processes $\eetoll$ ($l = e,\mu$)
are also studied.    
Figure \ref{fig:xwideinit} shows the MC distributions for two track 
final states which pass the $\pipi$, $\KK$ and $\ppbar$ criteria described above.  
The normalized center-of-mass energy variable $X_h$ is  
 defined as the sum of the energy of the two 
tracks (assuming the particle species of interest for each track) divided by 
$\sqrt{s}$.  
Figure \ref{fig:xwideinit} shows that the $\eetoKK$ and $\eetoppbar$ 
signal regions are sufficiently displaced from the dominant $\eetoll$ 
background, while the $\eetopipi$ signal overlaps with $\eetoll$.

\begin{table*}[!b]
\caption{Summary of form factor results.  The first errors are statistical only.  
The second errors in cross sections and form factors are systematic errors as 
discussed in the text.  The form factor results for protons correspond to the 
assumption $G^{p}_{E} = G^{P}_{M}$.  
Results for the assumption $G^{p}_{E} = 0$ are $\sim$9$\%$ larger.}
\begin{center}
\setlength{\tabcolsep}{5pt}
\hspace*{-0.6cm}\begin{tabular}{lccc}
\hline \hline
  & pion & kaon & proton\\
\hline
\# of Observed Events   & $26.0\pm5.1$ & $72.0\pm8.5$ & $16.0\pm4.8$\\
Lepton Contribution  & $\sim0.1$ & $0.6\pm0.2$  & $<0.1$\\
$\psi(2S)$ Contribution  & $<0.1$ & $0.6\pm0.1$ &  $1.9\pm0.2$  \\
\# of Net Signal Events  & $25.9\pm5.1$ & $70.9\pm8.5$ & $14.1\pm4.8$\\
$\sigma(e^+e^-\to h^+h^-)$, pb  & $9.0\pm1.8\pm1.3$ & $5.7\pm0.7\pm0.3$ 
& $1.2\pm0.4\pm0.1$\\
$F_h(|Q^2|)$ & $F_\pi$=0.075$\pm$0.008$\pm$0.005 & $F_K$=0.063$\pm$0.004$\pm$0.001 
& $G_M$=0.014$\pm$0.002$\pm$0.001\\
$Q^nF_h(|Q^2|)$, GeV$^2$  & $|Q^2|F_\pi$=1.01$\pm$0.11$\pm$0.07 
& $|Q^2|F_K$=0.85$\pm$0.05$\pm$0.02 & $|Q^4|G_M$=2.53$\pm$0.36$\pm$0.11\\
\hline \hline
\end{tabular}
\label{tab:results}
\end{center}
\end{table*}

In order to suppress $l^+l^-$ background events, it is first required that the 
accepted events have the ratio of the CC determined energy $E_{CC}$ and the 
track determined momentum $p$, $E_{CC}/p$, be less than 0.85.  For $p\bar{p}$ 
events this cut is only applied on the positive track, for $\pi^+\pi^-$ and 
$K^+K^-$ it is applied on both tracks.  In order to obtain a higher level of 
lepton rejection, signal to background optimization studies are made in terms 
of a likelihood variable defined by using RICH and $dE/dx$ information, 
$L(p,K)-L(l)=L_{RICH}(p,K)-L_{RICH}(l)+ \sigma^2_{dE/dx}(p,K)-\sigma_{dE/dx}^2(l)$. 
The optimum requirement is found to be $L(p,K)-L(e)<0$, and $L(p,K)-L(\mu)<-2$ for 
each track.

Rejecting leptonic background in the $\pi^+\pi^-$ sample requires additional 
measures.  These are determined by studying radiative Bhabha events in the 
continuum ($\sqrt{s}=3.671$ GeV) data, by studying $\mu$ tracks from the 
$e^+e^-\to\mu^+\mu^-$ Monte Carlo sample, and by studying  pions of appropriate 
momenta ($\sim1.83$ GeV/$c$)  in the existing CLEO sample of inclusive 
$D^0\to K^-\pi^+$ data taken at $\sqrt{s}=10.58$ GeV.  The optimization criteria 
which emerged from these studies are that pions must deposit $E_{CC}>0.42$ GeV, 
and  must have $L(\pi)-L(e)<0$, and $L(K)-L(\pi)>0$.  

Figure \ref{fig:xwidedata} 
shows the distribution of the events which meet all the above selection 
criteria as a function of $X_h$.  The signal region is defined as $0.98<X_h<1.02$ 
as bounded by the dashed lines.  The observed counts, the estimated lepton 
contamination counts, the counts contributed by the tail of the $\psi(2S)$ 
resonance, and the net signal counts are listed in Table \ref{tab:results}.  

\begin{table}[!b]
\caption{Sources of systematic uncertainty for the $\eetopipi$, $\eetoKK$,
and $\eetoppbar$ cross sections.}
\begin{center}
\begin{tabular}{lccc}
\hline
\hline
Source & $\pipi$ ($\%$) & $\KK$($\%$) & $\ppbar$ ($\%$) \\
\hline
Trigger & 1.0 & 1.0 & 1.0 \\
Tracking & 2$\times$1.0 & 2$\times$1.0 & 2$\times$1.0 \\
$X_{h}$ Signal Region & 4.1 & 0.5 & 3.7 \\
Net Momentum & 4.8 & 2.6 & 6.9 \\
$dE/dx$+RICH PID & 2$\times$2.7 & 2$\times$1.2 & 2$\times$1.6 \\
$E_{CC}$ & 10.7 & --- & --- \\
$\epsilon_{\pi}(E_{CC})$ & 2$\times$2.3 & --- & --- \\
MC statistics & 1.3 & 0.4 & 0.5 \\
$\psip$ Contamination & 0.08 & 0.1 & 1.0 \\
Leptonic Background & 0.05 & 0.3 & 0.0 \\
Radiative Correction & 0.2 & 0.2 & 0.2 \\
Luminosity, $\mathcal{L}$ & 1.0 & 1.0 & 1.0 \\
\hline
Total (in quadrature) & 14.6 & 4.4 & 8.9 \\
\hline
\hline
\end{tabular}
\end{center}
\label{tab:syst}
\end{table}

The net signal counts $N$ are related to the Born cross sections as
$\sigma_{0}(\eetohh) = N/[\epsilon{\cal{L}}(1+\delta)]$,
where $\epsilon$ is the detection efficiency, $\cal{L}$ is the 
total integrated luminosity, and 
$(1+\delta)$ is the radiative correction factor associated with $\hadpair$ 
production from $\ee$ annihilations.  
The proton and the kaon detection efficiencies, $\epsilon_p=0.657\pm0.003$, 
and $\epsilon_K=0.743\pm0.003$, are determined from the signal MC samples.  The 
pion detection efficiency, $\epsilon_\pi=0.166\pm0.002$, is determined by signal 
MC sample and the $D^0\to K^-\pi^+$ data.
The signal MC samples are generated with angular distributions according to 
Eqs. \ref{mesonad} and \ref{baryonad}.  
The initial state radiation corrections are determined using the method 
of Bonneau and Martin \cite{bonneaumartinrc} with the addition of 
$\mu$ and $\tau$ pair loops to the vacuum polarization term.  
For the final states $\pi^+\pi^-$, $K^+K^-$, and $p\bar{p}$, the $(1+\delta)$ 
correction factors are 0.832, 0.810, and 0.860, respectively.

The resulting cross sections are listed in Table \ref{tab:results}.  Various sources of 
systematic uncertainties in the cross sections have been studied.  These are 
listed in Table \ref{tab:syst}. Their sum in quadrature is 14.6\% for pions, 4.4\% for 
kaons, and 8.9\% for protons.  Integrating Eqs. \ref{mesonad} and 
\ref{baryonad} leads us to our final results 
for the form factors as listed in Table \ref{tab:results}~\cite{ffinterfere}.

The 2.89 pb$^{-1}$ of data taken at $\sqrt{s}=3.686$ GeV, or the $\psi(2S)$ 
resonance,  were used in the form factor analysis to evaluate the contribution 
that the resonance makes to the form factor data at $\sqrt{s}=3.671$ GeV.  For 
this purpose the $\psi(2S)$ data were analyzed in exactly the same way as the 
form factor data.  Our yields in the $\psi(2S)$ data sample, which we use to 
estimate the background from the $\psi(2S)$ tail feeding into the continuum 
sample, is consistent with the expectation based on the branching fractions 
\cite{pdg} for $p$ and $K$, but lower for $\pi$.

\begin{figure}[!tb]
\includegraphics*[width=3.6in]{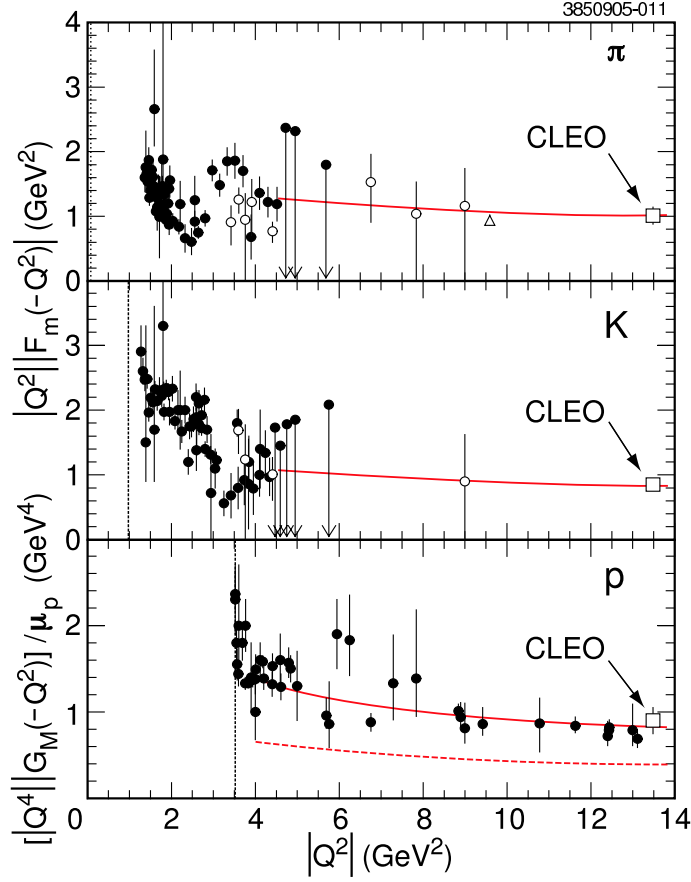}
\caption{Compilation of the existing experimental data for the 
pion (top), kaon (middle), and proton (bottom) form factors 
with timelike momentum transfer.  
Top and middle: The solid points are from identified $\pi^\pm$ and $K^\pm$ 
\cite{timelikepionkaon}. The open points are from unidentified $h^\pm$, divided 
into $\pi^\pm$ and $K^\pm$ according to VDM expectations.  For pions (top) the 
open triangle denotes the value obtained at $M^2(J/\psi)$ in Ref. \cite{jpsiff}.  
Bottom: The solid points for proton form factors are from the measurements and 
compilation of Ref. \cite{e760e835}.  The arbitrarily normalized solid curves 
show the variation of $\alpha_S(|Q^2|)$ (top and middle) and $\alpha_S^2(|Q^2|)$ 
(bottom), as determined for four flavors and $\Lambda=0.322$ GeV.  The dashed curve 
in the bottom plot shows $\alpha_S^2(|Q^2|)$ fit to the spacelike form factors of 
the proton.}
\label{fig:tlffcomb} 
\end{figure}

Our results for timelike form factors are displayed in 
Fig.~\ref{fig:tlffcomb} as $|Q^2|F_\pi$, $|Q^2|F_K$, 
and $|Q^4|G_M^p/\mu_p$, together with the existing world data for the same. 

Our precision result for the pion, $|Q^2|F_\pi(13.48~\mathrm{GeV}^2)$ = 
1.01$\pm$0.11$\pm$0.07 GeV$^2$, is the first such 
directly measured result.  It provides empirical validity for 
$|Q^2|F_\pi(9.6~\mathrm{GeV}^2)$ = (0.94$\pm$0.06) GeV$^2$ obtained by interpreting 
$\Gamma(J/\psi\to\pi^+\pi^-)/\Gamma(J/\psi\to e^+e^-)$ as a measure of the pion 
form factor \cite{jpsiff}.   Together, the two appear to support the pQCD 
prediction of $\alpha_S/|Q^2|$ variation of the form factor at large $|Q^2|$.  
Bebek {\sl et al.}\cite{bebek} have reported $|Q^2|F_\pi(9.77~\mathrm{GeV}^2)$ = 
0.69$\pm$0.19 GeV$^2$ for the spacelike form factor.  Within errors this is 
consistent with being nearly factor two smaller than the timelike form factors 
for $Q^2 > 9$ GeV$^2$, as found for protons.

Our measurement of the kaon form factor stands alone at present.  
The asymptotic pQCD prediction is \cite{pQCD} 
$F_{\pi,K}(|Q^2|)$ = $8\pi\alpha_{s}f^{2}_{\pi^+,K^+}/|Q^2|$.  
For $\alpha_{s}$ = 0.3, $f_{\pi^+}$ = $130.7\pm0.4$ MeV, and 
$f_{K^+}$ = $159.8\pm1.5$ MeV \cite{pdg}, this leads to 
$F_\pi(13.48~\mathrm{GeV}^2)$ = 0.010 and $F_K(13.48~\mathrm{GeV}^2)$ = 0.014, 
which are factors of $\sim$8 and $\sim$4 smaller than our results, respectively.  
The $\alpha_s$ and $Q^2$ independent pQCD prediction $F_\pi(|Q^2|)/F_K(|Q^2|)$ 
= $f_\pi^2/f_K^2$ = 0.67$\pm$0.01 is also in disagreement with our result 
$F_\pi(13.48\;\mathrm{GeV}^2)/F_K(13.48~\mathrm{GeV}^2)$ = 1.19$\pm$0.17.  
Bakulev {\sl et al.}\cite{softcontributions} have estimated soft contributions 
to the pion form factor in the framework of QCD sum rules.  Their formulation 
leads to $F_\pi(13.48~\mathrm{GeV}^2)$ = 0.010 and 
$F_\pi(|Q^2|)/F_K(|Q^2|)$ = 0.51, i.e., the discrepancy between our experimental 
results and the theoretical predictions remains.  We note that this behavior 
is in contrast to the good agreement between the measured $\pi^{0}\gamma$ 
transition form factor and the pQCD prediction for the same \cite{CLEOpi0}.
 
%$F_\pi(|Q^2|)/F_K(|Q^2|)$ \cite{pQCD}.  With the pQCD prediction is 
%$F_\pi(|Q^2|)/F_K(|Q^2|)=0.67\pm0.01$. Our measured result, 
%$F_\pi(13.48\;\mathrm{GeV}^2)/F_K(13.48\;\mathrm{GeV}^2)=1.19\pm0.17$, is in 
%disagreement with this prediction.

Our result for $G^p_M(13.48~\mathrm{GeV}^2)$ is in excellent agreement with the 
results of the Fermilab E760/E835 experiments in which the reverse reaction 
$p\bar{p}\to e^+e^-$ was measured \cite{e760e835}. Our results provide the 
first independent confirmation of the Fermilab results.

We gratefully acknowledge the effort of the CESR staff in providing us with 
excellent luminosity and running conditions. This work was supported by the 
National Science Foundation and the U.S. Department of Energy.


\begin{thebibliography}{99}
\bibitem{ffinterfere} The present form factor analysis has been done without 
taking into account possible interference between continuum and $\psi(2S)$ cross 
sections.  For the extremes of completely constructive and completely 
destructive interference it is estimated that the form factors for $\pi$, $K$, 
and $p$ change by $\pm3.5\%$, $\pm4.5\%$, and $\pm5.9\%$, respectively.
\end{thebibliography}
\end{document}